\documentclass{epl}

\newcommand{\pdag}{{\phantom{\dagger}}}
\newcommand{\bq}{\begin{equation}}
\newcommand{\eq}{\end{equation}}
\newcommand{\bn}{\begin{eqnarray}}
\newcommand{\en}{\end{eqnarray}}

\title{Mesoscopic Kondo effect of a quantum dot embedded in an Aharonov-Bohm ring with %%@
intradot spin-flip scattering}
\author{Bing Dong\inst{1,2} \and G. H. Ding\inst{2} \and H. L. Cui\inst{1,3} \and X. L. %%@
Lei\inst{2}}
\institute{
  \inst{1} Department of Physics and Engineering Physics, Stevens Institute of %%@
Technology, Hoboken, New Jersey 07030, USA\\
  \inst{2} Department of Physics, Shanghai Jiaotong University, 1954 Huashan Road, %%@
Shanghai 200030, China \\
  \inst{3} School of Optoelectronics Information Science and Technology, Yantai %%@
University, Yantai, Shandong, China
}

\begin{document}

\maketitle

\begin{abstract}

We study the Kondo effect in a quantum dot embedded in a mesoscopic ring taking into %%@
account intradot spin-flip scattering $R$. Based on the finite-$U$ slave-boson %%@
mean-field approach, we find that the Kondo peak in the density of states is split into %%@
two peaks by this coherent spin-flip transition, which is responsible for some %%@
interesting features of the Kondo-assisted persistent current circulating the ring: (1) %%@
strong suppression and crossover to a sine function form with increasing $R$; (2) %%@
appearance of a ``hump" in the $R$-dependent behavior for odd parity. $R$-induced %%@
reverse of the persistent current direction is also observed for odd parity.

\end{abstract}

Intensive attention has been paid to the Kondo effect in a quantum dot %%@
(QD),\cite{Goldhaber,Cronenwett} in which formation of a bound state between the local %%@
spin and conduction electrons induces a Kondo peak in the density of states (DOS) at the %%@
Fermi level $E_{\rm F}$, consequently leading to strong enhancement of the linear %%@
conductance ultimately to the ideal value $2e^2/h$ below the bulk Kondo temperature %%@
$T_{\rm K}^0$. Associated with this energy scale, the characteristic length of the bound %%@
state wave-function, the so-called Kondo screening cloud defined as $\xi_{\rm %%@
K}^0\approx \hbar v_{\rm F}/T_{\rm K}^0$ ($v_{\rm F}$ is the Fermi velocity), is itself %%@
mesoscopic size (of the order of $1~\mu$m). 

The easy control of major parameters of these artificial atoms in a wide range makes it %%@
possible to study the Kondo effect even when the size of the lead is comparable to the %%@
size of screening cloud $\xi_{\rm K}^0$. Recently, Affleck and Simon\cite{Affleck} %%@
suggested that the setup of a QD embedded in a closed mesoscopic Aharonov-Bohm (AB) ring %%@
is an ideal tool to detect this cloud because their perturbation calculations found that %%@
the persistent current (PC) in such a geometry is a universal scaling function of %%@
$\xi_{\rm K}^0/L$ ($L$ is the ring circumference), and crosses over from the saw-tooth %%@
of perfect ring in the limit $\xi_{\rm K}^0/L\ll 1$ to a sine function with vanishing %%@
magnitude at $\xi_{\rm K}^0/L\gg 1$. Besides, slave-boson mean-field (SBMF) evaluation %%@
disclosed that this finite size effect affects drastically the Kondo resonance in a way %%@
that depending on the total number of electrons ({\em modelo} $4$).\cite{Hu,Ding} There %%@
was also a variational calculation about this problem.\cite{Kang}  

On the other hand, the recent issue has been focused on what happens to the Kondo %%@
physics if an intrinsic spin-flip scattering is stirred in the QD and/or even when the %%@
conduction band is allowed itself to be ferromagnetic %%@
metal.\cite{Rudzinski,Souza,Zhang,Dong,Ma,Lopez,Choi} By means of the SBMF approach, Ma %%@
{\it et al}.\cite{Ma} and L\'opez {\it et al}.\cite{Lopez} found respectively a %%@
splitting of the zero-bias peak of the nonlinear differential conductance when the %%@
spin-flip scattering amplitude $R$ is of the order of the Kondo temperature, indicating %%@
the Kondo peak splitting and suppression in the DOS. This has been further confirmed by %%@
using an exact theoretical method, numerical renormalization group technique.\cite{Choi} 

Actually, the electron in the semiconductor QD suffers inevitably intrinsic relaxations %%@
(decoherence) due to the spin-orbital interaction\cite{Fedichkin} or the %%@
hyperfine-mediated spin-flip transition,\cite{Erlingsson} which can be modeled by a %%@
phenomenological spin-flip scattering term $R$ as done in these previous %%@
papers.\cite{Ma,Lopez,Choi} Naturally, it would be interesting to study the modification %%@
of the Kondo resonance due to the intradot spin-flip transition at a mesoscopic %%@
situation $\xi_{\rm K}^0/L\sim 1$ and its effect on the PC. 

In this letter, based on the finite-$U$ SBMF approach initially developed by Kotliar and %%@
Ruckenstein,\cite{KR} we perform a detailed analysis for this issue. With the help of %%@
four slave boson parameters, this SBMF approach not only allows to deal with finite %%@
on-site Coulomb interaction but also is expected to give good results both for the %%@
charge fluctuation and spin fluctuation of the Anderson impurity model, which enables us %%@
to thoroughly analyze this problem (the Kondo-type DOS in QD without attaching to leads %%@
and the PC circulating the ring) sweeping across the whole regions, i.e., Kondo, %%@
mixed-valence, and empty-orbital regimes.\cite{Ding,Ma}  

We consider a QD with a single bare level $\epsilon_{d}$ having a finite on-site Coulomb %%@
repulsion $U$ and an intradot spin-flip scattering $R$ is embedded in a mesoscopic AB %%@
ring, modeled by a one-dimensional tight-binding Hamiltonian with the nearest neighbor %%@
hopping $t$, through tunneling $t_{\eta}$ ($\eta=L,R$). According to the finite-$U$ %%@
slave-boson approach,\cite{KR} we introduce four auxiliary boson operators $e$, %%@
$p_{\sigma}$, and $d$, which are associated respectively with the empty, singly occupied %%@
and doubly occupied electron states at the QD, to describe the above physical problem %%@
without correlation terms in an enlarged space with constraints. In terms of the %%@
auxiliary particle representation, the effective Hamiltonian for this system (totally %%@
$N$ sites including the QD) can be written as:   
\begin{eqnarray}
H&=&-t\sum_{j=1}^{N-2}\sum_\sigma(c_{j\sigma}^{\dagger} c_{j+1\sigma}^{\pdag} + {\rm %%@
H.c.}) + \sum_{\sigma} \epsilon_d
c_{d\sigma}^{\dagger} c_{d\sigma}^{\pdag} \cr
&& + Ud^{\dagger} d + Rc_{d\uparrow}^{\dagger} c_{d\downarrow}^{\pdag} + R %%@
c_{d\downarrow}^{\dagger} c_{d\uparrow}^{\pdag}  - \sum_{\sigma}(t_L %%@
z_{\sigma}^{\dagger} c_{d\sigma}^{\dagger} c_{1\sigma}^{\pdag} + t_R %%@
e^{i\phi}c_{N-1\sigma}^{\dagger} c_{d\sigma}^{\pdag} z_{\sigma}^{\pdag} + {\rm H.c.}) %%@
\cr
&& + \lambda^{(1)}(\sum_{\sigma}p_{\sigma}^{\dagger} p_{\sigma}^{\pdag} + e^{\dagger} e %%@
+ d^{\dagger} d - 1 )  + \sum_{\sigma}\lambda_{\sigma}^{(2)}(c_{d\sigma}^{\dagger} %%@
c_{d\sigma}^{\pdag} -p_{\sigma}^{\dagger} p_{\sigma}^{\pdag} - d^{\dagger} d ), %%@
\label{h}
\end{eqnarray}
where $c_{d\sigma}^{\dagger}$ ($c_{d\sigma}$) is the creation (annihilation) operators %%@
of electrons in the QD, the three Lagrange multipliers $\lambda^{(1)}$ and %%@
$\lambda_{\sigma}^{(2)}$ impose the constraints, and in the hopping term the fermion %%@
operator $c_{d\sigma}$ is replaced by $c_{d\sigma} z_{\sigma}$ with a many-body %%@
correction factor $z_{\sigma}= (1-d^\dagger d-p_\sigma^\dagger %%@
p_\sigma^{\pdag})^{-\frac{1}{2}} ( e^\dagger p_\sigma^{\pdag} + p_{\bar %%@
{\sigma}}^\dagger d ) (1-e^\dagger e-p_{\bar {\sigma}}^\dagger
p_{\bar {\sigma}}^{\pdag} )^{-\frac{1}{2}}$. The phase factor $\phi$ is defined by $2\pi %%@
\Phi/\Phi_{0}$, in which $\Phi$ and $\Phi_0=h/e$ are the external magnetic flux enclosed %%@
by the ring and the flux quantum, respectively. The coupling strength $\Gamma$ between %%@
the QD and the ring is associated with the ring DOS $\rho(E_{\rm F})$ and hopping %%@
$t(E_{\rm F})$ at the Fermi energy as $\pi \rho(E_{\rm F}) |t(E_{\rm F})|^2$. Here we %%@
would like to point out that the intrinsic spin-flip processes lift the level degeneracy %%@
of the QD, yielding $\epsilon_d \pm R$. Because of no special spin direction, it seems %%@
that our model is equivalent to the one of the QD under external magnetic field (Zeeman %%@
effect). However, we still prefer the present form since it provides a direct %%@
description of spin relaxation in the QD and can give us a heuristic comprehension of %%@
the spin-relaxation effects on the PC, although the spin is flipped without any type of %%@
dissipation in this simple model. Moreover, our results obtained in the following are %%@
also applicable for the case of applying external magnetic field. We consider the %%@
half-filled case ($N_{e}=N$) in this letter, and in the continuum limit we have $E_{\rm %%@
F}=0$, $\rho(0)=N/(2\pi t)$, $|t(0)|^2=2(t_L^2+t_R^2)/N$, and $\Gamma_{\eta}=t_\eta^2/t$ %%@
($\eta=L,R$).

At zero temperature, we utilize the mean-field approximation in which all the boson %%@
operators are replaced by real {\em c} numbers (variational parameters). Consequently we %%@
can solve the noninteracting Hamiltonian Eq.~(\ref{h}) exactly by numerical %%@
diagonalization and evaluate the expectation value of any operators in the ground state %%@
$|0\rangle$ by summing over consecutive eigenlevels up to the highest occupied single %%@
particle energy level (HOEL). Minimizing the ground state energy $E_{\rm g}$ %%@
[expectation value of the Hamiltonian Eq.~(\ref{h}) in the ground state] with respect to %%@
these variational parameters together with the three constraints form the basic %%@
self-consistent equations to determine these unknown parameters within the mean-field %%@
scheme.\cite{Ding} Once solving the set of nonlinear equations, one can calculate the %%@
electron DOS $\rho_{\sigma}(\omega)$ for the QD by
\bq
\rho_{\sigma}(\omega)=-\frac{1}{\pi} |z_{\sigma}|^2 {\rm Im} \langle v| c_{d %%@
\sigma}^{\pdag} (\omega+i 0^{+} - H)^{-1} c_{d \sigma}^{\dagger} |v\rangle,
\eq
($|v\rangle$ stands for the vaccum state) and the PC $I$ by
\bq
I=-\frac{e}{\hbar} \frac{\partial E_{\rm g}}{\partial \phi} = \frac{ie}{\hbar} %%@
\sum_{\sigma} \langle 0| t_{R} z_{\sigma} e^{i \phi} c_{N-1 \sigma}^{\dagger} %%@
c_{d\sigma}^{\pdag}-{\rm H.c.} |0\rangle.
\eq
The mesoscopic ``Kondo correlation energy" $T_{\rm K}$ is defined as the binding energy %%@
between the QD and the ring:
\bq
T_{\rm K}=\epsilon_{d}-\epsilon_{\rm F}-(E_{\rm g}-E_{\rm g}^0), \label{Kondo}
\eq
in which $\epsilon_{\rm F}$ is the energy of the HOEL corresponding to the Fermi energy %%@
and $E_{\rm g}^0$ is the ground state energy of the $N-1$ sites tight-binding %%@
Hamiltonian with the same electron number $N_{e}$.   

In the following, we present our calculations only for the mesoscopic ring with %%@
$\xi_{\rm K}^0/L\sim 1$ ($L=N$) in the particle-hole symmetric region. We will also take %%@
$t=1$, $t_L=t_R=0.35$, $\Gamma=\Gamma_L+\Gamma_R=0.245$ and choose the on-site %%@
interaction $U=8\Gamma$, for the symmetric (asymmetric) case $\epsilon_d=-U/2$ ($-U/4$) %%@
corresponding to $T_{\rm K}^0=0.013$ ($0.034$) or $\xi_{\rm K}^0=148$ ($59$). 

\begin{figure}
\twofigures[width=7cm,height=6.5cm,angle=0,clip=on] {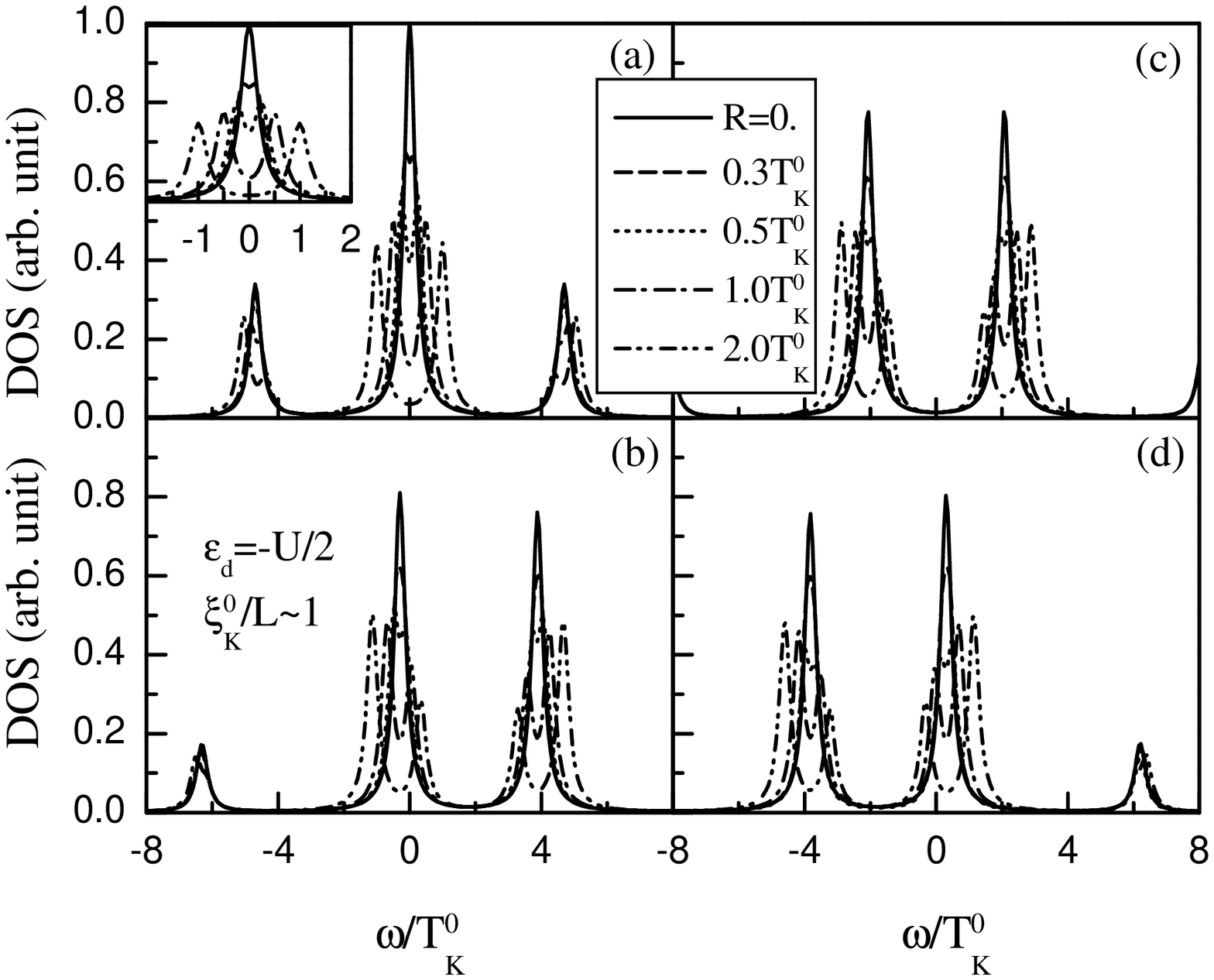}{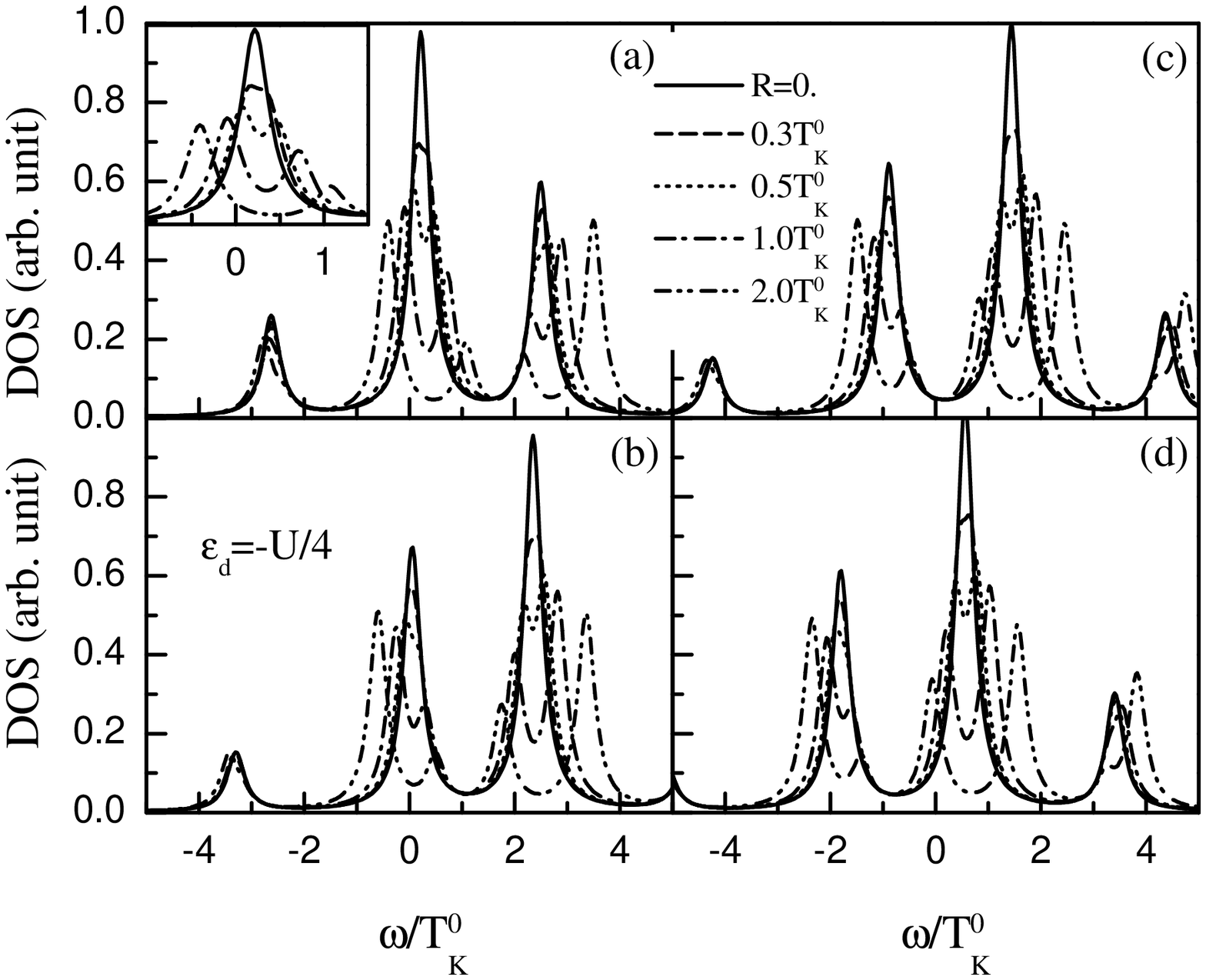}
\caption{The dot DOS for the symmetric case $\epsilon_{d}=-U/2$ at various spin-flip %%@
processes $R$ without magnetic flux. The number of the lattice sites $N$ are (a) $4n$, %%@
(b) $4n+1$, (c) $4n+2$, and (d) $4n+3$. Inset: $\rho_{\sigma}(\omega)$ around $\omega=0$ %%@
at $N=4n$.} 
\label{fig1}
\caption{The dot DOS $\rho(\omega)$ for the asymmetric case $\epsilon_{d}=-U/4$. The %%@
other parameters are the same as in Fig.~1.}
\label{fig2}
\end{figure}

Figure 1 shows the dot DOS $\rho_{\sigma}(\omega)$ in the symmetric case %%@
$\epsilon_{d}=-U/2$ for different values of the intradot spin-flip scattering $R$ in %%@
terms of the Kondo temperature $T_{\rm K}^0=U\sqrt{\beta}\exp(-\pi/\beta)/2\pi$ %%@
[$\beta=-2U\Gamma/\epsilon_{d}(U+\epsilon_{d})$]. It is obvious that in the mesoscopic %%@
case $L \sim \xi_{\rm K}^0$, the bulk single Kondo peak becomes a set of subpeaks %%@
depending on the total numbers of lattice sites $N$ (modulo $4$), which are roughly %%@
attributed to the patterns of the HOEL and lowest unoccupied energy level (LUEL) of the %%@
Hamiltonian Eq.~(\ref{h}).\cite{Hu} For $N=4n$, the energies of the HOEL ($\epsilon_{\rm %%@
F}$) and the LUEL are both nearly equal to zero, so there is a single main peak in DOS %%@
pinned at $\omega=0$ together with several small peaks located at other single particle %%@
energy levels. For $N=4n+2$, $\epsilon_{\rm F}$ is negative with a distinguishable value %%@
and LUEL is equal to $-\epsilon_{\rm F}$. As a result, two Kondo peaks appear around %%@
$\omega=0$ with the same amplitudes. While in the cases of odd parity ($N=4n\pm 1$), %%@
$\epsilon_{\rm F}$ is nearly below or above zero, respectively, leading to a single %%@
Kondo resonance slightly shifted from the Fermi energy. This shifting of the Kondo peak %%@
from the Fermi energy $\epsilon_{\rm F}$ in the odd parity will cause the reduction of %%@
PC magnitude (see Fig.~4 below). 
On the other hand, introduction of the spin-flip scattering $R$ always reduces and %%@
splits every Kondo peak regardless of parity. Moreover, with increasing $R$ the DOS %%@
suppresses in a more pronounced way, and the two split Kondo peaks are entirely %%@
symmetric for $N=4n$. However, the situation is somewhat different for the asymmetric %%@
system $\epsilon_{d}=-U/4$ as shown in Fig.~2. The profiles of DOS wholly move a little %%@
bit towards the high energy. This is because the asymmetric system allows charge %%@
fluctuations to certain extent and consequently increases the resulting single particle %%@
energy levels delicately. Associated with the spin-flip induced splitting, this feature %%@
results in a novel $R$-dependent behavior of the PC (see below). 

We also study the splitting width $\delta$ of the main Kondo peak at $N=4n$ and plotted %%@
it as a function of $R$ in Fig.~3(a), which shows a significant reduction of the width %%@
$\delta$ in contrast to the predicted value $2R$ by measuring nonlinear differential %%@
conductance through a QD connected to macroscopic leads.\cite{Lopez} Actually, the exact %%@
Bethe-ansatz solution for the QD also reveal explicitly this large suppression of the %%@
peak splitting due to the strong correlations in the presence of the external Zeeman %%@
field.\cite{MW} Besides, Fig.~3(b) displays that, for the fixed $R$, the peak splitting %%@
shrinks when the QD departs away from the particle-hole symmetric point. In Fig.~3(c) %%@
and Fig.~3(d), we summarize the calculated mesoscopic Kondo temperatures $T_{\rm K}$ %%@
with variation of $R$ by using Eq.~(\ref{Kondo}). We find that: (1) The finite size %%@
effect remarkably increases the binding energy between the QD and the ring, so the %%@
mesoscopic Kondo temperatures $T_{\rm K}$ are much higher than the corresponding bulk %%@
values $T_{\rm K}^0$; (2) Charge fluctuations can decrease this binding energy (because %%@
it increases the single particle energy levels), and accordingly, lead to a smaller %%@
$T_{\rm K}$ in the particle-hole asymmetric region; (3) The behaviors of $T_{\rm K}$ %%@
with rising of $R$ depend on the parity: For even parity $N=4n\pm 2$, $T_{\rm K}$ %%@
decreases first with small amplitude and then begins to increase, while in the case of %%@
odd parity $N=4n\pm 1$, $T_{\rm K}$ increases monotonously, implying that the Kondo %%@
screening cloud shrinks due to the spin-flip scattering.      

\begin{figure}[t]
\includegraphics [width=8cm,height=6.5cm,angle=0,clip=on] {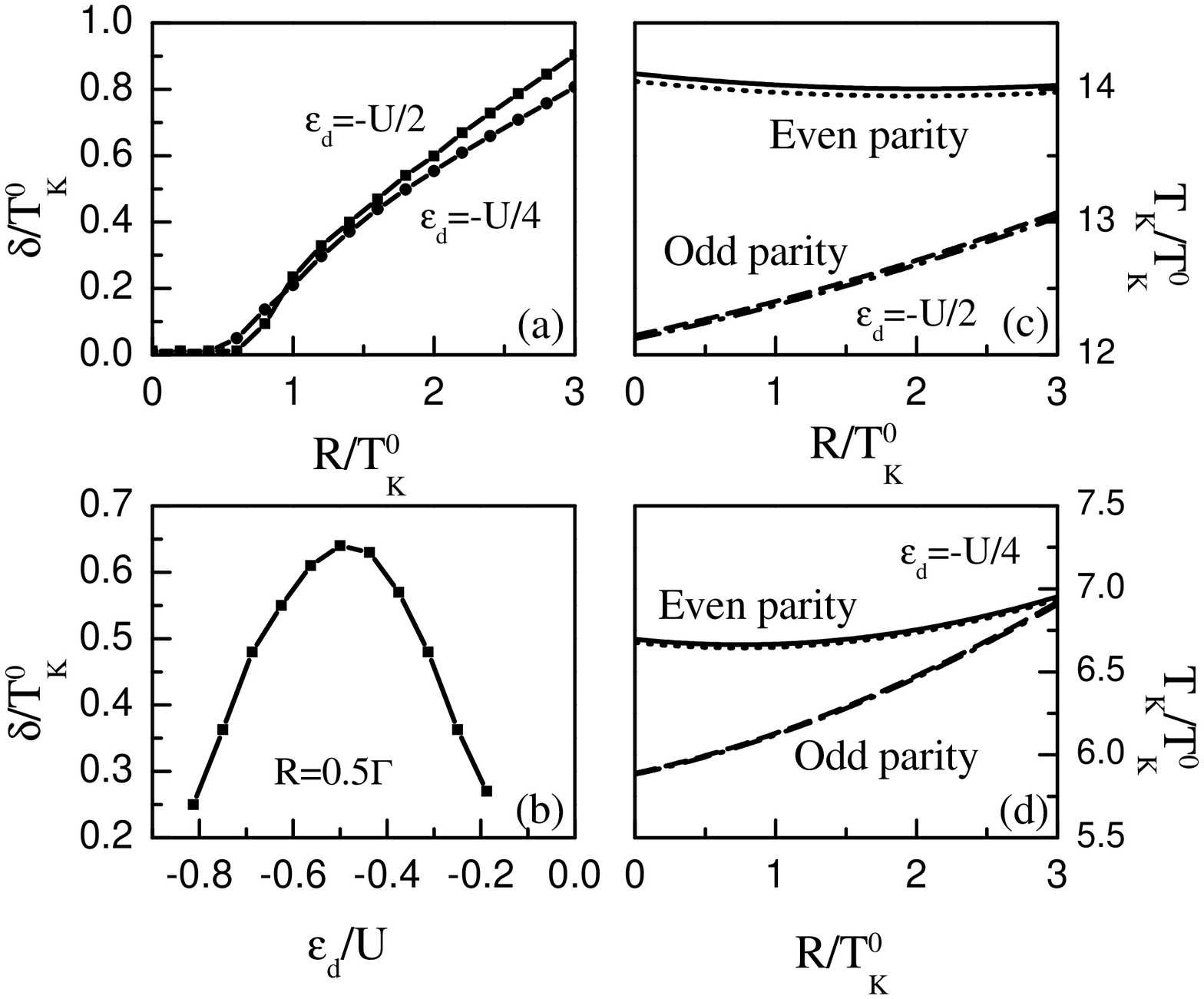}
\caption{(a) Splitting $\delta$ of the main Kondo peak vs. $R$ for $\epsilon_{d}=-U/2$ %%@
and $-U/4$ at $N=4n$ and zero magnetic flux. (b) $\delta$ as a function of %%@
$\epsilon_{d}$ at $R=0.5 \Gamma$. Kondo temperature $T_{\rm K}/T_{\rm K}^0$ vs. $R$ for %%@
(c) the symmetric and (d) the asymmetric systems at mesoscopic condition $L\sim \xi_{\rm %%@
K}^0$ and $\phi=\pi/2$.} \label{fig3}
\end{figure}

\begin{figure}[htb]
\makebox{\includegraphics[width=4.3cm,height=4.5cm]{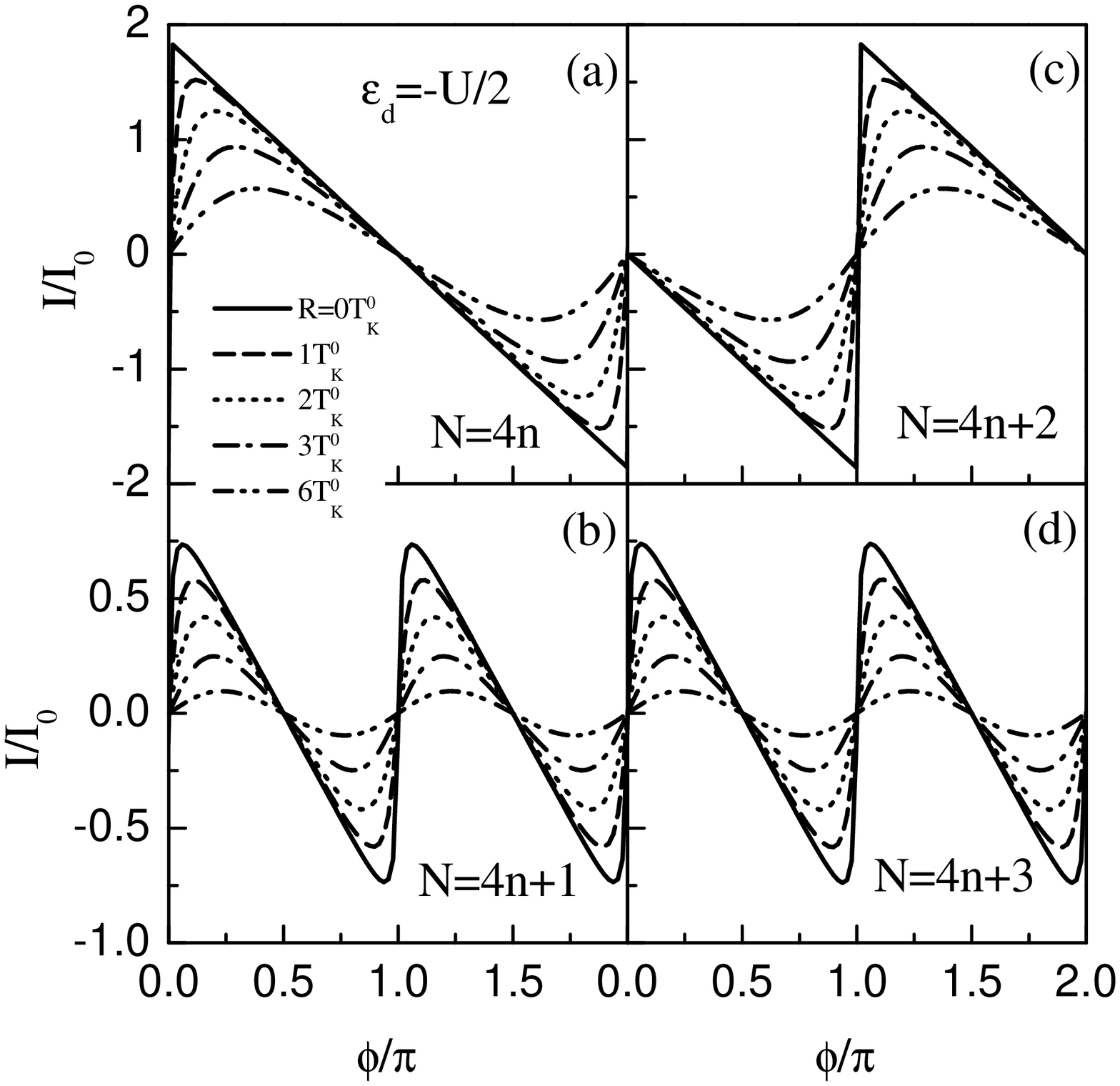}
		 \includegraphics[width=4.3cm,height=4.5cm]{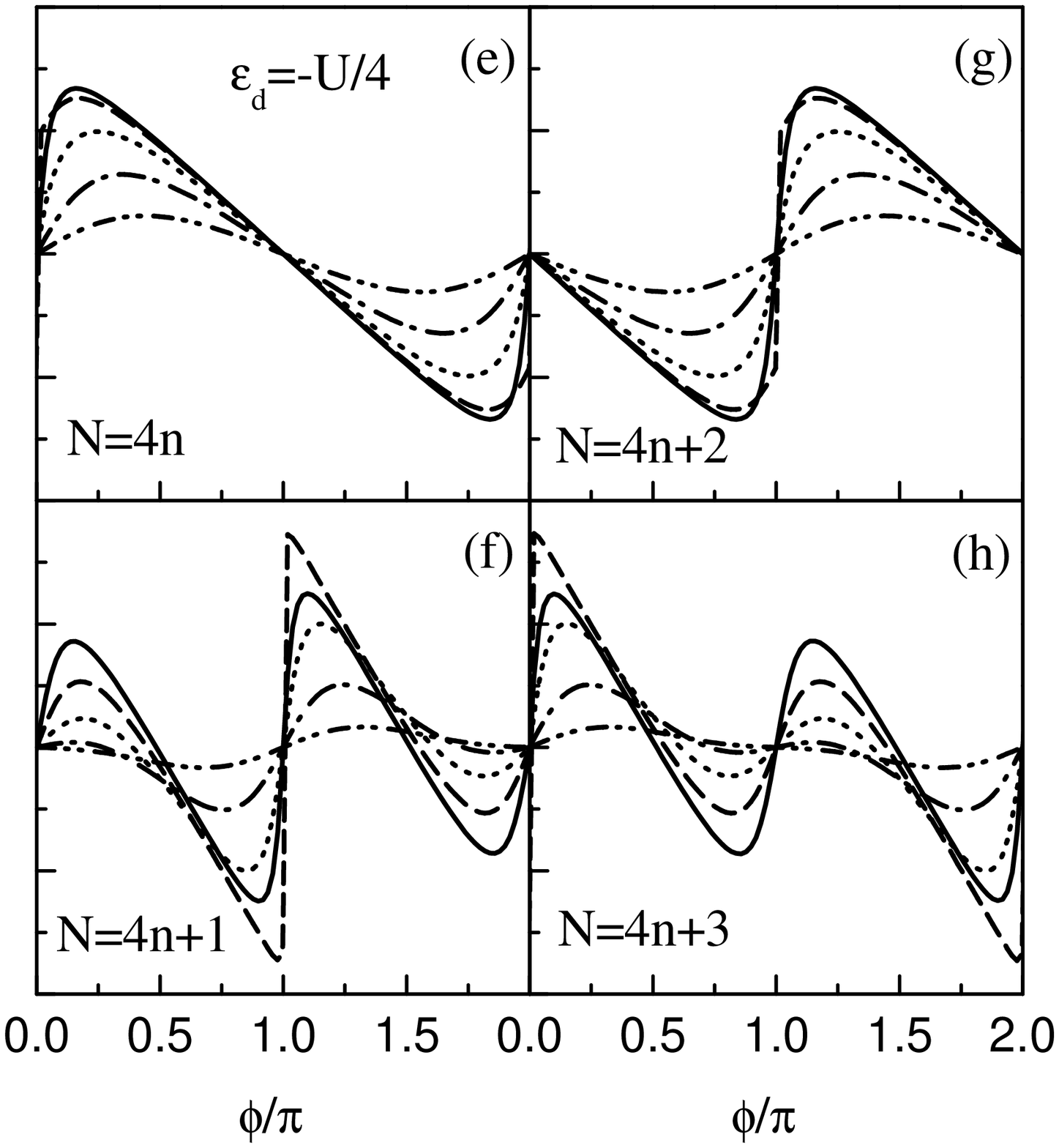} }
\caption{Normalized PCs vs. magnetic flux at various $R$ for $\epsilon_{d}=-U/2$ (a)-(d) %%@
and $\epsilon_{d}=-U/4$ (e)-(h). $I_0=2t/N$ is the corresponding PC of an ideal $N$ %%@
sites ring.} \label{fig4}
\end{figure}

Now we turn to calculation of the PC circulating around the ring, which is believed to %%@
be a useful way to detect the Kondo resonance due to the fact that the PC is closely %%@
related to electron transmission through the QD embedded in the ring. We illustrate in %%@
Fig.~4 the $R$-dependent PC $I/I_0$ vs. magnetic flux at mesoscopic condition for %%@
$\epsilon_{d}=-U/2$ and $\epsilon_{d}=-U/4$. In the absence of spin-flip transition, (1) %%@
the PC has an analogous structure of an ideal ring, implying high transmission through %%@
the QD; (2) its direction depends on the parity of the sites number $N$; (3) the %%@
magnitude of the PC for the odd parity is much smaller than that for the even parity %%@
because of the Kondo peak slightly shifting in the case of the odd parity; and (4) the %%@
perfect symmetric structures for the particle-hole symmetric point $\epsilon_{d}=-U/2$ %%@
guarantee $I^{4n+1}(\phi)=I^{4n-1}(\phi)$, but there are no this feature for the %%@
asymmetric systems. As expected, applying spin-flip scattering generally lowers the %%@
magnitude of the PC due to weakening of the Kondo resonance and varies its profile %%@
gradually to a rough sinusoidal, even though the screening cloud is also reduced at the %%@
same time. More strikingly, strong spin-flip scattering even changes the sign of %%@
$dI^{4n+1}(\phi)/d\phi$ at small $\phi$ for the asymmetric system from positive to %%@
negative, indicating that the symmetry of the HOEL changes and the system becomes %%@
diamagnetic.      

\begin{figure}[t]
\twofigures[width=7cm,height=6.5cm,angle=0,clip=on] {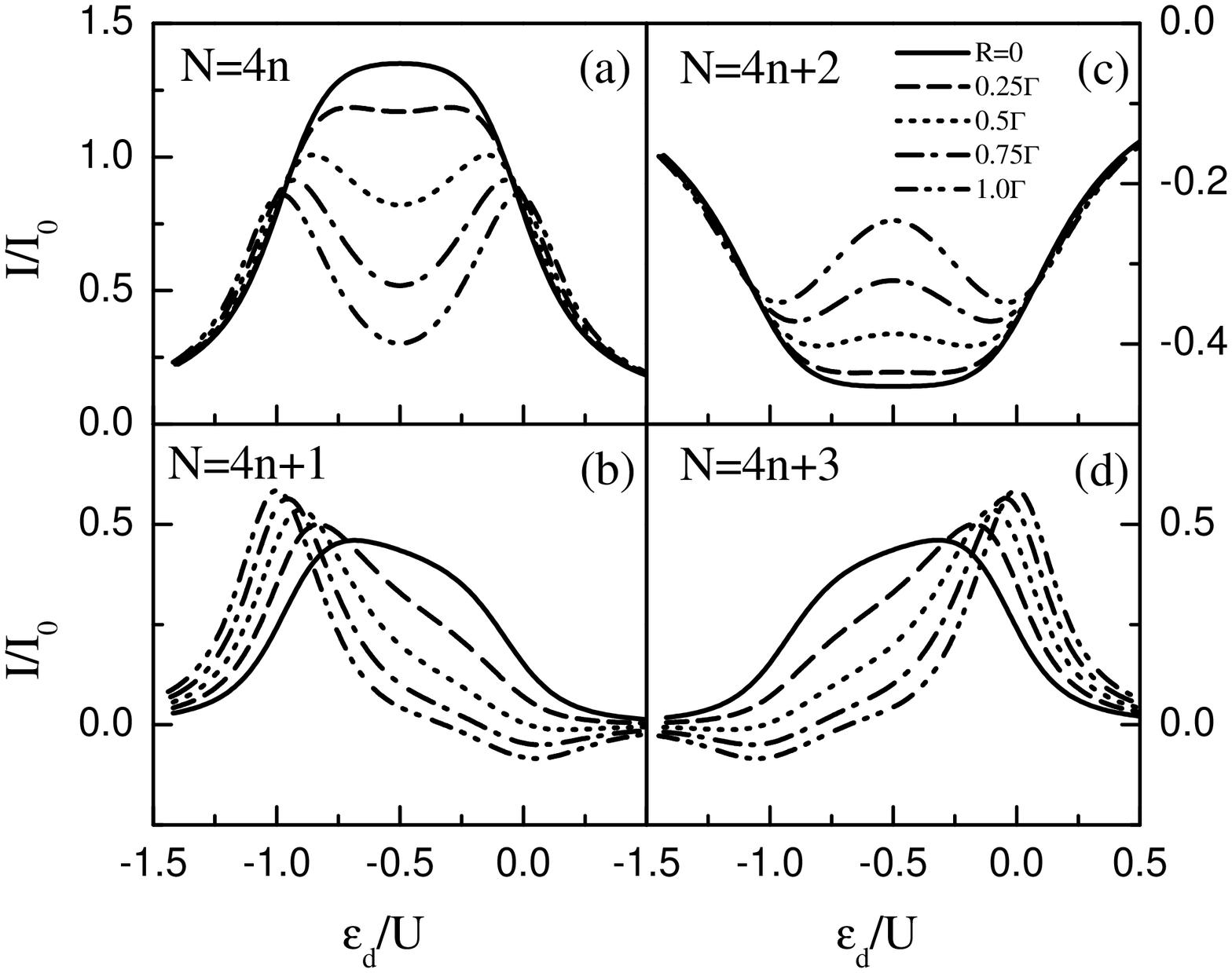}{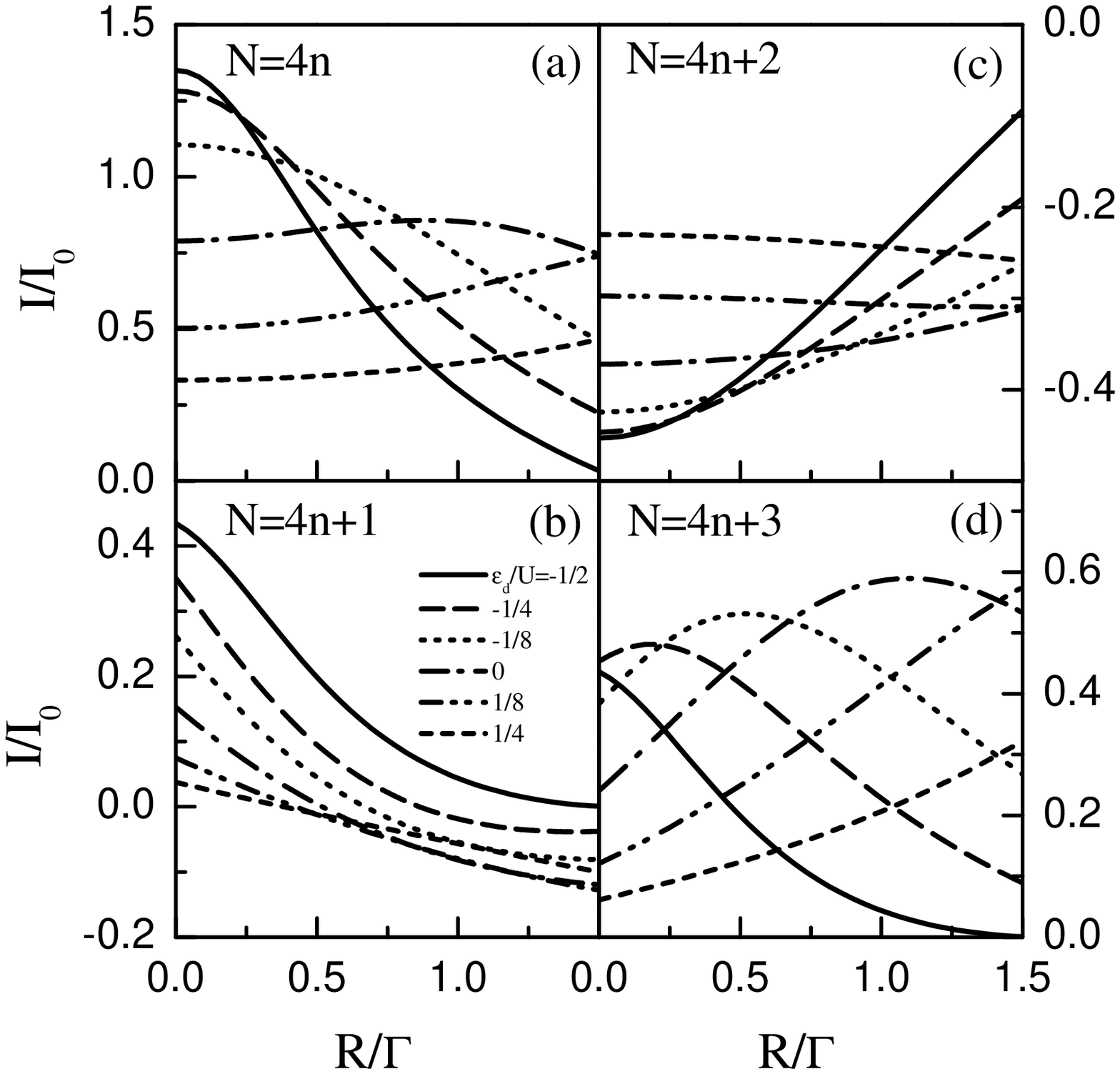}
\caption{PC vs. energy level $\epsilon_{d}/U$ of the QD for different parities and %%@
several spin-flip transitions at $\phi=\pi/4$.} 
\label{fig5}
\caption{The PC as a function of spin-flip scattering $R$ at different energy levels of %%@
the QD.}
\label{fig6}
\end{figure}

Figure 5 displays the PC as a function of the QD level. For even parities, the PC %%@
exhibits symmetric feature around the central point $\epsilon_{d}=-U/2$ and unique %%@
magnetic response properties (being paramagnetic for $N=4n$, while diamagnetic for %%@
$N=4n+2$) in the whole range of energy levels. In the region where the Kondo effect %%@
plays a dominating role, the PC is obviously enhanced in contrast to its corresponding %%@
ideal value ($I/I_0>1$). However, significant suppression of the Kondo resonance due to %%@
increasing $R$ can reduce the Kondo-assisted PC rapidly. But in both non-Kondo and deep %%@
Kondo regimes, we observe this reduction gets gradually weak and even a slight increase %%@
because of charge fluctuations. Finally at a large scattering $R=1.0\Gamma$, the PC %%@
shows a Coulomb blockade pattern in analogy with the behavior of linear conductance %%@
through QD over the Kondo temperature. For the case of odd parity, the situation is %%@
quite different. The Kondo-assisted PC is asymmetric and a remarkable feature can be %%@
clearly seen $I^{4n+1}(\epsilon_d/U)=I^{4n+3}(|\epsilon_{d}/U|-1)$ due to the symmetry %%@
properties of DOS in the finite-$U$ Anderson model. Two more striking results can be %%@
observed (we only address for $N=4n+3$ here): (1) The PC is a monotonous function of the %%@
spin-flip $R$ at $\epsilon_{d}>0$ and $\epsilon_{d}<-U/2$, whereas there is no universal %%@
$R$ behavior in the middle widow; (2) As mentioned above, the PC may change direction %%@
with increase of $R$ for $\epsilon_{d}<-U/2$. We summarize these $R$-dependent features %%@
of the PC more clearly in Fig.~6, in which we plot the PC as a function of $R$. %%@
Explicitly in Fig.~6(d), the PC exhibits a ``hump" structure for %%@
$\epsilon_{d}/U=-\frac{1}{4}$ and $-\frac{1}{8}$ as increasing $R$, which is due to the %%@
fact that the Kondo resonance is shifted away from the Fermi energy as shown in %%@
Fig.~2(d).

In short, we have investigated the Kondo resonance in a QD embedded in a mesoscopic AB %%@
ring taking into account intradot spin-flip scattering $R$ or application of magnetic %%@
field in the QD by means of the finite-$U$ SBMF approach. The Kondo resonance splits %%@
into two clear peaks (Zeeman effect) with a largely reduced width when increasing $R$. %%@
We also reported the $R$-dependent PC behaviors in both Kondo and non-Kondo regimes, %%@
which can measure this Kondo effect in an equilibrium condition. This model is %%@
experimentally accessible for the update techniques and provides a possible test for the %%@
influence of spin-flip transition on the Kondo screening.    
       
\acknowledgments

BD and HLC is supported by the DURINT Program administered by the US Army Research %%@
Office. XLL is supported by Major Projects of National Natural Science Foundation of %%@
China, the Special Founds for Major State Basic Research Project and the Shanghai %%@
Municipal Commission of Science and Technology.

\end{document}